\documentclass[final,twocolumn]{elsarticle}

\usepackage{amssymb}
\usepackage{amsmath}

\usepackage{graphicx}
\usepackage{dcolumn}
\usepackage{bm}
\usepackage{xcolor}
\usepackage{multirow}
\usepackage{makecell}
\usepackage{placeins}
\usepackage{soul}
\usepackage{lineno}
\usepackage{xurl}
\usepackage{hyperref}
\usepackage{ulem}
\usepackage{setspace}
\usepackage{cellspace}

\newcommand{\DL}[1]{\textcolor{blue}{#1}} 

\journal{Solid State Physics}

\begin{document}

\begin{frontmatter}



\title{Higher order magnetoelasticity energy corrections in bcc and fcc systems} 


\author[VSB]{Jakub Šebesta} 
\ead{jakub.sebesta@vsb.cz}

\author[UK]{Ondřej Faiman}%

%

\author[UK,VSB]{Dominik Legut}
\ead{dominik.legut@matfyz.cuni.cz}

\affiliation[VSB]{organization={IT4Innovations, VSB – Technical University of Ostrava},
            addressline={17. listopadu 2172/15}, 
            city={Ostrava},
            postcode={708 00},
            country={Czech Republic}}

\affiliation[UK]{organization={Charles University, Faculty of Mathematics and Physics},
            addressline={Ke Karlovu 3}, 
            city={Praha 2},
            postcode={121 16},
            country={Czech Republic}}

\begin{abstract}
Magnetoelastic properties play a vital role in industrial applications. Despite being hidden behind either purely magnetic or elastic behavior, magnetoelasticity takes place in a wide range of devices as transducers, acoustic actuators, or fast response sensors. In this work, we inspect the impact of higher-order terms on the anisotropic magnetostriction behavior.  Regarding ab-initio calculations, the anisotropic magnetostriction can be related to the strain dependence of the magnetocrystaline energy.  Commonly, the description is restricted to a linear strain dependence in the magnetoelastic energy. Here, we derive higher-order terms in strain for bcc and fcc crystal {structures}. Using a simple parametrization, we show that the influence of the higher-order strain terms is negligible for the studied {cubic} systems.
\end{abstract}



\begin{keyword}
magnetoelasticity, magnetostriction, magnetism  


\end{keyword}

\end{frontmatter}




\section{Introduction}

Regarding magnetic materials,  mechanical properties become more complex as there arises an interplay between elastic and magnetic properties -- so-called magnetoelasticity~\cite{chikazumi2009physics}. The relation between magnetism and mechanical properties offers a wide range of applications, including rotational and linear motors, force sensors, or low-energy demanding electronics employing magneto-acoustics\cite{Spetzler_scirep_r21,Calkins_jimss_r07,Ekreem_jmpt_r07,Bienkovsky_SensAct_r04,Kuszewski_iop_r18}. The most evident magnetoelastic effects are volume and axial--anisotropic magnetostrictions. The first represents a change of the volume of a magnetized sample irrespective of magnetization direction, whereas the second denotes magnetization direction dependent change of sample dimensions ~\cite{MAELAS_1_r21}.

In theory, the anisotropic magnetostriction can be related to a strain dependence of magnetocrystalline anisotropic energy (MAE) with respect to the magnetization direction \cite{Nieves_sss_r25_seconorder}. 
It offers a feasible way to estimate the magnetostriction behavior in the framework of ab-initio calculation~\cite{MAELAS_1_r21,MAELAS_2_r22}. Commonly, the approximation of magnetoelastic energy considers only the linear strain term as it is used in ab-initio estimation of magnetostriction parameters~\cite{MAELAS_1_r21,MAELAS_2_r22} or numerical simulations describing experimentally observed features, e.g., an interaction of surface acoustic waves with spin waves~\cite{r21_Babu_SAW_SW,r25_Ngouagnia_SAW_SW}.

Since an occurrence of discrepancies between the ab-initio and experimentally estimated anisotropic magnetoelastic parameters {has been reported}~\cite{MAELAS_2_r22}, here, we inspect a possible impact of higher order terms in the magnetoelastic energy description. Expanding it to higher order strain contributions, we focus on the modification of the theoretical anisotropic magnetostriction behavior in order to explain the "plain" correspondence with the experiment. Namely, we discuss cubic bcc and fcc systems commonly used for the description of Fe, resp. Ni.

\section{Theory}

The anisotropic magnetostriction denotes changes in the sample dimension arising from its magnetization. Unlike the volume magnetostriction, it depends on the magnetization direction. A relative change of the length along direction $\bm{\beta}$ when a demagnetized system is magnetized in  $\bm{\alpha}$ direction follows \cite{MAELAS_1_r21,cullen_book}
\begin{equation}
   \frac{l-l_{0}}{l} \bigg{\vert}^{\bm{\alpha}}_{\bm{\beta}}  =  \sum_{i,j=x,y,z} \varepsilon^{{eq}}_{ij} {\beta}_{i} {\beta}_{j} \, ,
   \label{Eq.magstrictcoeff}
\end{equation}
where $l_{0}$ and $l$ are the  length of demagnetized resp. magnetized sample, and $\varepsilon^{{eq}}_{ij}$ denotes equilibrium strain tensor.
Assuming small lattice deformations, the strain tensor reads 
\begin{equation}
    \varepsilon_{ij} = \left( \frac{\partial u_{i}}{\partial r_{j}} + \frac{\partial u_{j}}{\partial r_{i}} \right) \, ,
    \label{Eq:strain_tensor}
\end{equation}
where  $\mathbf{u}(r)=\mathbf{r}^{\prime}-\mathbf{r}$ describes a displacement vector.
The equilibrium strain $\varepsilon^{{eq}}_{ij}$ is related to minimization of  magnetoelastic energy $E_{\mathrm{me}}$ and elastic energy $E_{\mathrm{el}}$ contributions with a strain as follows~\cite{engdahl2000handbook,chikazumi2009physics,CLARK1980531}
\begin{equation}
    \frac{\partial  (E_{\mathrm{me}} + E_{\mathrm{el}})}{\partial \varepsilon^{{eq}}_{ij}} = 0  \, . \label{Eq:Eminimiz}
\end{equation}.

Concerning the cubic symmetry, the elastic energy in Cartesian coordinates follows~\cite{MAELAS_1_r21,AELAS_r17}
\begin{align}
    &\frac{1}{V_{0}}( E_{\mathrm{el}}-E_{0}) =    \quad
   \frac{1}{2}c_{\mathrm{xxxx}}(\varepsilon_{\mathrm{xx}}^{2}  + \varepsilon_{\mathrm{yy}}^{2} + \varepsilon_{\mathrm{zz}}^{2}) \nonumber \\
    &+ c_{\mathrm{xxyy}}(\varepsilon_{\mathrm{xx}}  \varepsilon_{\mathrm{yy}}+\varepsilon_{\mathrm{xx}}  \varepsilon_{\mathrm{zz}}+ \varepsilon_{\mathrm{yy}}  \varepsilon_{\mathrm{zz}}) \nonumber \\
    &
    + 2 c_{\mathrm{yzyz}}(\varepsilon_{\mathrm{xy}}^{2} +\varepsilon_{\mathrm{yz}}^{2} + \varepsilon_{\mathrm{zx}}^{2}) \nonumber  \\
    &=\frac{1}{2}C_{11}(\varepsilon_{\mathrm{1}}^{2} + \varepsilon_{\mathrm{2}}^{2}+ \varepsilon_{\mathrm{3}}^{2}) + C_{12}(\varepsilon_{\mathrm{1}}\varepsilon_{\mathrm{2}} + \varepsilon_{\mathrm{1}}\varepsilon_{\mathrm{3}} + \varepsilon_{\mathrm{2}}\varepsilon_{\mathrm{3}}) \nonumber \\
    & +\frac{1}{2} C_{44}(\varepsilon_{\mathrm{4}}^{2} + \varepsilon_{\mathrm{5}}^{2} + \varepsilon_{\mathrm{6}}^{2})   \, ,
    \label{Eq.Eel_cubic}
\end{align}
where $E_{0}$ and $V_{0}$ denote the equilibrium energy and volume, and  $c_{ijkl}$ resp. $C_{ij}$ represent elastic constants. 

The generally used magnetoelastic (ME) energy linear in the strain reads~\cite{MAELAS_1_r21,MAELAS_2_r22}
\begin{align}
\label{Eq:MagelEng}
   \frac{1}{V_{0}} E_{me} &=b_{0}(\varepsilon_{\mathrm{xx}}  + \varepsilon_{\mathrm{yy}} + \varepsilon_{\mathrm{zz}}) \nonumber \\
    & + b_{1}(\alpha_{x}^{2}\varepsilon_{\mathrm{xx}} + \alpha_{y}^{2}\varepsilon_{\mathrm{yy}} + \alpha_{z}^{2}\varepsilon_{\mathrm{zz}})  \nonumber \\    &+ 2 b_{2} (\alpha_{x} \alpha_{y} \varepsilon_{\mathrm{xy}} + \alpha_{x} \alpha_{z} \varepsilon_{\mathrm{xz}}  + + \alpha_{y} \alpha_{z} \varepsilon_{\mathrm{yz}} )  \nonumber \\
   & = b_{0}(\varepsilon_{\mathrm{1}}  + \varepsilon_{\mathrm{2}} + \varepsilon_{\mathrm{3}})  + b_{1}(\alpha_{x}^{2}\varepsilon_{\mathrm{1}} + \alpha_{y}^{2}\varepsilon_{\mathrm{2}} + \alpha_{z}^{2}\varepsilon_{\mathrm{3}}) \nonumber \\    &+  b_{2} (\alpha_{x} \alpha_{y} \varepsilon_{\mathrm{6}} + \alpha_{x} \alpha_{z} \varepsilon_{\mathrm{5}}  + + \alpha_{y} \alpha_{z} \varepsilon_{\mathrm{4}} ) \, .
\end{align}
where $b_{0}$ is the isotropic magnetoelastic constant and $b_{1}$, $b_{2}$  are the anisotropic ones. The formula can be derived from the dipole-dipole interaction under strain. Assuming collinear spin structures, the interaction follows
\begin{equation}
    E_{dd} = l(r) \Big( \bm{\alpha}\cdot\tilde{\bm{\beta}} -\frac{1}{3} \Big) , 
    \label{Eq:ddinteracion}
\end{equation}
where $l(r)$ denotes a coupling depending on the bond length $r$ and $\tilde{\bm{\beta}}$ is the bond direction.

Minimizing the sum of the elastic and magnetoelastic energy (Eq.~\ref{Eq:Eminimiz}), the  equilibrium strain reads
\begin{align}
   &\varepsilon^{{eq}}_{ii}= -\frac{b_{1}\alpha_{i}^{2}}{C_{11}-C_{12}}  -\frac{b_{0}}{C_{11}+2C_{12}} \nonumber  \\  &+ \frac{b_{1}C_{12}}{(C_{11}-C_{12})(C_{11}+2C_{12})} , \label{Eq.equilb_strain_ii} \\
   & \varepsilon^{{eq}}_{ij}= -\frac{b_{2}\alpha_{i}\alpha_{j}}{2C_{44}} , i\neq j \;  \label{Eq.equilb_strain_ij}.
\end{align}
Assuming the Eq.~\ref{Eq.magstrictcoeff} the relative change of the length reads%
\begin{align}
   &\frac{l-l_{0}}{l} \bigg{\vert}^{\bm{\alpha}}_{\bm{\beta}}  = \lambda^{\alpha} + \frac{3}{2}\lambda_{001}(  \alpha_{x}^{2} {\beta}_{x} + \alpha_{y}^{2} {\beta}_{y} + \alpha_{z}^{2} {\beta}_{z}  - \frac{1}{3})  \label{Eq.lambda001111}  \\
   &+ 3 \lambda_{111}(\alpha_{x}\alpha_{y}\beta_{x}\beta_{y} + \alpha_{y}\alpha_{z}\beta_{y}\beta_{z} + \alpha_{x}\alpha_{z}\beta_{x}\beta_{z}) \, \nonumber ; \\
  & \lambda^{\alpha} = \frac{-b_{0}-b_{1}/3}{C_{11}+2C_{12}} , \lambda_{001}=\frac{-2b_{1}}{3(C_{11}-C_{12})}, \lambda_{111}= \frac{-b_{2}}{3C_{44}} \nonumber
\end{align}
where $\lambda^{\alpha}$, $\lambda_{001}$ and  $\lambda_{111}$ denote isotropic resp. anisotropic magnetostriction coefficients.

\FloatBarrier

\section{Results}

The commonly used definition of the magnetoelastic energy (Eq.~\ref{Eq:MagelEng}) is based on a simple expansion in the strain of the dipole-dipole interaction (Eq.~\ref{Eq:ddinteracion}), including only the first-order strain terms. Here, we do a more detailed expansion of the magnetoelastic energy in bcc and fcc structures, considering higher-order terms. 
A~bond vector $\mathbf{r}$ between neighboring atoms changes  under strain as follows
\begin{equation}
\mathbf{r}^{\prime}=
\mathbf{r}_{0} +\bm{\varepsilon}\mathbf{r}_{0}
\, ,
\label{Eq:strainded_r}
\end{equation}
where $\bm{\varepsilon}$ is a symmetric strain tensor (Eq.~\ref{Eq:strain_tensor}) and $\mathbf{r}_{0}$, $\mathbf{r}^{\prime}$ denote original resp. strained vector.
 Thereby, the  bond length $r$ in the coupling constant $l(r)$ and the bond direction $\bm{\beta}$ representing direction cosines are modified by the strain 
(Eq.~\ref{Eq:ddinteracion}).

 To include higher order strain terms to magnetoelastic energy, we expand the strain in terms of the changes in $r$ and $\tilde{\bm{\beta}}$. Considering Eq.~\ref{Eq:strainded_r}, a strained bond length $r^{\prime}$ reads 
\begin{align}
    r^{\prime} =\vert \mathbf{r^{\prime}} \vert = r_{0} \;  \Bigg\vert 1+\frac{\bm{\varepsilon}\mathbf{r}_{0}}{r_{0}^2} \Bigg\vert \; ,
\end{align}
where $r_{0}$ denotes unstrained bond length. To obtain high-order strain terms,  the norm is expanded employing the Taylor series of ${\sqrt{1+x}}$. Here we consider expansion up to the third order ${\sqrt{1+x}} \sim 1 +\frac{x}{2} - \frac{x^2}{8} +\frac{x^3}{16} + \mathcal{O}(x^4)$.

Similarly, a Cartesian component of the strained direction cosine $\beta_{i}^{\prime}$ reads
\begin{equation}
    \tilde{\beta}_{i}^{\prime} = \frac{r_{i}^{\prime}}{ r_{0} } \frac{1}{\vert 1+(\bm{\varepsilon}\mathbf{r}_{0})/r_{0}^2 \vert} \; ,
    \label{Eq.:dircoschange}
\end{equation}
where one can applied Taylor series of $\frac{1}{\sqrt{1+x}} \sim 1 -\frac{x}{2} + \frac{3x^2}{8} -\frac{5x^3}{16} + \mathcal{O}(x^4)$ to expand the strain dependence in the fraction.  Separating strain given contributions to bond length $\Delta r$ and direction cosine $\Delta \tilde{\bm{\beta}}$
\begin{align}
    &r^\prime = r + \Delta r  \; , \label{Eq.raddist_mod} \\
    &\tilde{\beta}_{i}^{\prime} = \tilde{\beta}_{i} + \Delta \tilde{\beta}_{i} \; , \label{Eq.dircosmod}
\end{align}
the dipole-dipole interaction 
(Eq.~\ref{Eq:ddinteracion})  under strain

 \begin{align}
   &E_{dd}(\bm{\varepsilon}) = \Big( l(r)  + \frac{\partial l}{\partial r} \Big\vert_{r_{0}}  \Delta {r}\Big) \Bigg[  (\bm{\alpha}\cdot\tilde{\bm{\beta}})^{2} +  (\bm{\alpha}\cdot\Delta\tilde{\bm{\beta}})^{2}   - \frac{1}{3}  \Bigg] \nonumber \\
    & = l(r_{0}) \Bigg[  (\bm{\alpha}\tilde{\bm{\beta}})^{2} - \frac{1}{3} \Bigg] + \frac{\partial l}{\partial r}\Big\vert_{r_{0}}  \Delta {r} \Bigg[ (\bm{\alpha}\cdot\tilde{\bm{\beta}})^{2}  - \frac{1}{3} \Bigg]\nonumber \\ 
    &+   l(r_{0})(\bm{\alpha}\cdot\Delta\tilde{\bm{\beta}})^{2} + \frac{\partial l}{\partial r} \Big\vert_{r_{0}} \Delta {r}  \Bigg[  (\bm{\alpha}\cdot\Delta\tilde{\bm{\beta}})^{2}   \Bigg] \nonumber \\
    &  = l(r_{0}) \Bigg[  (\bm{\alpha}\tilde{\bm{\beta}})^{2} - \frac{1}{3} \Bigg] + \Delta E_{dd} \label{Eq:Edd_difference}
     \; .
\end{align}
Summing the strain-dependent $\Delta E_{dd}$ terms (Eq.~\ref{Eq:Edd_difference}) over the neighbors in the crystal lattice, the form of the magnetoelastic energy $E_{me}$ can be derived. In this work, one restricts only to the nearest-neighbor atoms. A detailed derivation of the magnetoelastic energy in the case of bcc and fcc structure{s} is shown in the {\ref{APPENDIX}}.

Performing the summation over the nearest neighbors,  the anisotropic part of the magnetoelastic energy  up to the second order in the strain follows a general form
\begin{align}
  &  \frac{1}{V_{0}} E_{me}^{\mathrm{II}} =    
      b_{1} \sum_{i} \varepsilon_{ii}  \alpha_{i}^{2}    + {b_{2}} \sum_{i\neq j} \varepsilon_{ij}  \alpha_{i} \alpha_{j} 
 + {{b_{1}^{\prime}} }  \sum_{i\neq j}  \varepsilon_{ii}    \alpha_{j}{^{2}} \nonumber \\
 & + b_{1}^{\prime\prime}  \sum_{i} \varepsilon_{ii}^{2}  \alpha_{i}^{2}
 +  b_{2}^{\prime\prime} \sum_{i\neq j}  \varepsilon_{ii}^{2}    \alpha_{j}{^{2}}
  +   b_{3}^{\prime\prime} \sum_{i\neq j} \varepsilon_{ij}^{2} \alpha_{i}^{2}  \nonumber  \\
 & + b_{4}^{\prime\prime} \sum_{i\neq k,j\neq k }  \varepsilon_{ij}^{2}    \alpha_{k}{^{2}} 
  +  b_{5}^{\prime\prime} \sum_{i\neq j}   \varepsilon_{ii} \varepsilon_{jj}  \alpha_{i}^{2} \nonumber \\
    & +    b_{6}^{\prime\prime} \sum_{i\neq k,j\neq k } \varepsilon_{ii} \varepsilon_{jj} \alpha_{k}^{2}   
 +  b_{7}^{\prime\prime} \sum_{i\neq j}  \varepsilon_{ii} \varepsilon_{ij}  \alpha_{i} \alpha_{j} \nonumber \\
    &
   + b_{8}^{\prime\prime} \sum_{i\neq k,j\neq k } \varepsilon_{ij} \varepsilon_{kk}\alpha_{i} \alpha_{j} 
   +  b_{9}^{\prime\prime} \sum_{i\neq k,j\neq k } \varepsilon_{ik} \varepsilon_{jk}\alpha_{i} \alpha_{j} 
   \label{Eq:2nd_ME}
\end{align}

where  $b_{i}^{\prime}$ resp. $b_{i}^{\prime\prime}$ denotes anisotropic magnetoelastic constant related to extra terms to the cubic magnetoelastic energy (compare Eqs.~\ref{Eq:MagelEng} and~\ref{Eq:2nd_ME}). Higher-order terms are shown in the 
{\ref{APPENDIX}}.
Following the derivation,  the magnetoelastic constants $b$ (Eq.~\ref{Eq:2nd_ME}) are parametrized by $l(r)$ and $r_{0}\frac{\partial l}{\partial r}\Big\vert_{r_{0}}$ values, where the actual form depends on the symmetry of the considered structure (Table~\ref{Tab:magel_constants}).

\begin{table}[h]
\centering
\caption{\label{Tab:magel_constants} Parametric expressions for the  magnetoelastic constant in terms of $l=l(r)$ and  $L= r_{0} \frac{\partial l}{\partial r} \Big\vert_{r_{0}}$.  }%
\setstretch{1.5}

\begin{tabular}{ccc}
\hline
\hline
 \phantom{-------}&  bcc & fcc   \\
\hline
$b_{1}$ & $\frac{32}{9} l$ & $2(2l+L) $  \\
$b_{2}$ & $\frac{16}{9}( l +  L ) $ & $2(2l+L)$ \\
$b_{1}^{\prime}$ & $-\frac{16}{9} l $ & $(-2l+L) $  \\
$b_{1}^{\prime\prime}$ & $-\frac{16}{27}(l  -2 L)$  &  $ -\frac{1}{2}(4 l  -5 L) $\\
$b_{2}^{\prime\prime}$ & $\frac{8}{27}(l  -2 L) $ &  $ \frac{1}{4}(4 l  -3 L) $ \\
$b_{3}^{\prime\prime}$ & $-\frac{8}{27}(l  -4 L)$ &  $  \frac{1}{2}(2 l + L) $\\
$b_{4}^{\prime\prime}$ & $\frac{8}{27}(l  -4 L)$ &  $ -(2 l  -L) $\\
$b_{5}^{\prime\prime}$ & $-\frac{16}{27}(l  -L)$ & $ -\frac{1}{2}L $\\
$b_{6}^{\prime\prime}$ & $\frac{16}{27}(2 l  -L)$ & 0 \\
$b_{7}^{\prime\prime}$ & $-\frac{80}{27}(l  -L) $ &  $2 L $\\
$b_{8}^{\prime\prime}$ & $\frac{32}{27}(l  -L)$ &    $ -4(2 l  -L)  $\\
$b_{9}^{\prime\prime}$ & $ \frac{8}{27}(11   -5 L)$ & $ -2(6 l  -5 L)  $\\
\hline
\hline
\end{tabular}

\end{table}

\begin{figure}[h]
    \centering
    \includegraphics[width=\linewidth]{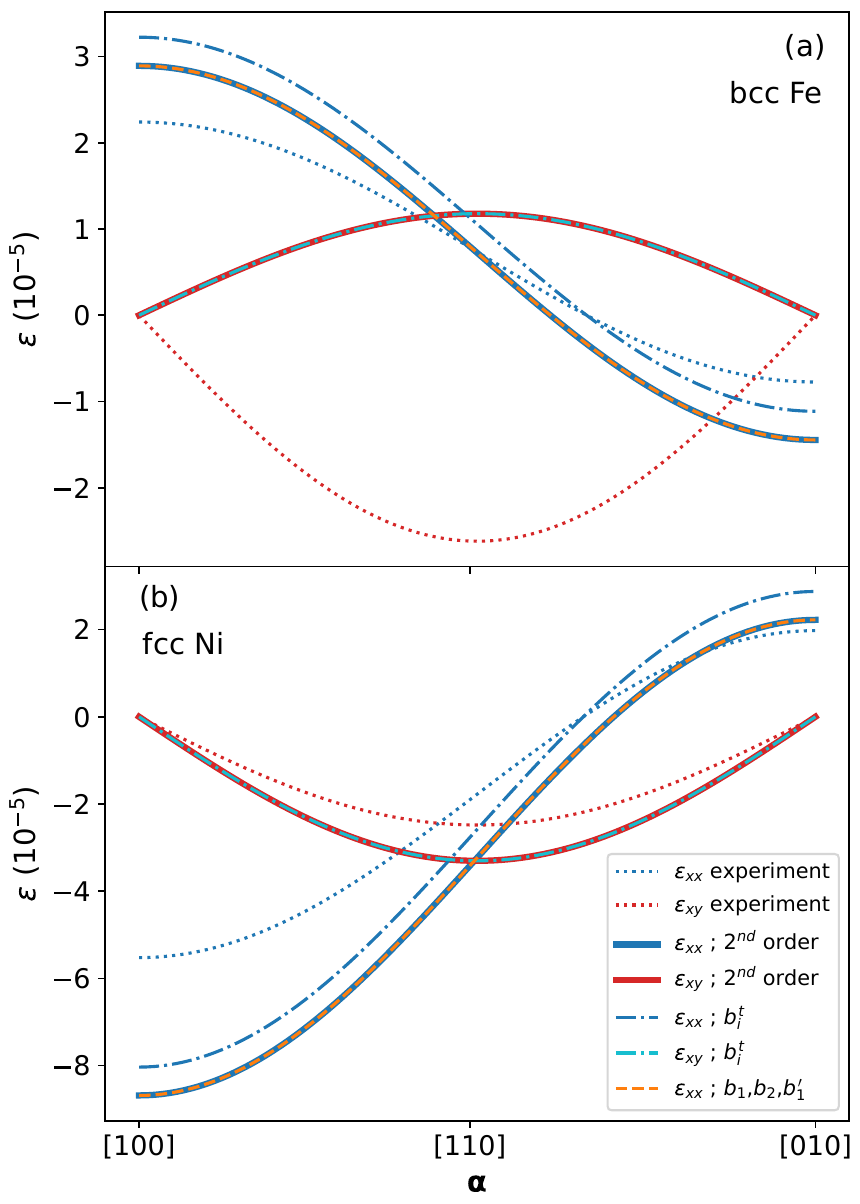}
    \caption{Equilibrium strain as a function of the magnetization direction. (a) bcc Fe (b) fcc Ni. (solid lines) Equilibrium strain with the ME energy up to 2$^{nd}$ order in the strain (Eq.~\ref{Eq:2nd_ME}). (dotted lines resp. dash-dotted lines) Behavior for experimental constants $b_{1}^{e}$ and $b_{2}^{e}$ resp. ab-initio one $b_{1}^{t}$ and $b_{2}^{t}$  with original ME energy formula (Eq.~\ref{Eq:MagelEng}). (dotted lines) Equilibrium strain with only the linear strain terms in the Eq.~\ref{Eq:2nd_ME}.  }
    \label{fig:bcc_fcc}
\end{figure}

Unlike  the original magnetoelastic energy (Eq.~\ref{Eq:MagelEng}), in the case of high-order corrections (Eq.~\ref{Eq:2nd_ME}), it is non-trivial to find a general solution for the equilibrium strain (Eq.~\ref{Eq:Eminimiz}) and to define $\lambda$ parameters (Eq.~\ref{Eq.lambda001111})  due to a set of strongly coupled equations with a high number of $b$ parameters (compare Eqs.~\ref{Eq:mimim_set_orig_1}-\ref{Eq:mimim_set_orig_6} and Eqs.~\ref{Eq:minimiz_full1},~\ref{Eq:minimiz_full4}). Therefore, to analyze the influence of the extended formula of the magnetoelastic energy, we evaluate the equilibrium strains for particular elements, considering bcc Fe and fcc Ni systems, see    (Fig.~\ref{fig:bcc_fcc}).

Here, we do not perform ab-initio calculations of the magnetoelastic and elastic constants, but consider already published  values~\cite{MAELAS_2_r22,MAELAS_1_r21}. Since only the linear magnetoelastic constants $b_{1}$ and $b_{2}$ have been fitted, we model the values of the other constants based on the parametrization (Table~\ref{Tab:magel_constants}) arising from the expansion of the dipole-dipole interaction under strain. Regarding both structures, bcc and fcc, values of $l=l(r)$ and  $L= r_{0} \frac{\partial l}{\partial r}$ can be expressed in terms of constant\DL{s} $b_{1}$ and $b_{2}$.  
However, there exists a difference in definition between the ab-initio determined magnetoelastic constants $b_{1}^{t}$~\cite{MAELAS_2_r22} and constants $b_{1}$ in the Eq.~\ref{Eq:2nd_ME}.
According to the Eq.~\ref{Eq:MagelEng}, $b_{1}^{t}$  was estimated from slope of the strain dependence of the total energy difference $E(\alpha_{1},\varepsilon_{xx}) - E(\alpha_{2},{\varepsilon_{xx}})$,  where $\alpha_{1}=[100]$ and $\alpha_{2}=[\frac{1}{\sqrt{2}  }\frac{1}{\sqrt{2}} 0]$ are two distinct magnetization directions. Assuming the original magnetoelastic energy  Eq.~\ref{Eq:MagelEng}, it yields
\begin{equation}
    \frac{1}{V_{0}} \Big[E_{me}([100],\varepsilon_{xx}) - E_{me}([\frac{1}{\sqrt{2}  }\frac{1}{\sqrt{2}} 0],\varepsilon_{xx}) \Big]= \frac{1}{2}b_{1}\varepsilon_{\mathrm{xx}} \, .
\end{equation}
Whereas, concerning the magnetoelastic energy up to the second order (Eq.~\ref{Eq:2nd_ME}), the energy difference reads
\begin{align}
    \frac{1}{V_{0}} E_{me}^{\mathrm{II}}([100],\varepsilon_{xx}) - E_{me}^{\mathrm{II}}([\frac{1}{\sqrt{2}  }\frac{1}{\sqrt{2}} 0],\varepsilon_{xx}) \nonumber \\ = \frac{1}{2}(b_{1}\varepsilon_{\mathrm{xx}}-b_{1}^{\prime}\varepsilon_{\mathrm{xx}}+b_{1}^{\prime\prime}\varepsilon_{\mathrm{xx}}^{2} - b_{2}^{\prime\prime}\varepsilon_{\mathrm{xx}}^{2} ) \, .
    \label{Eq.b1b3_ediff}
\end{align}
Therefore, neglecting the quadratic terms (Eq.~\ref{Eq.b1b3_ediff}) in our model, it is assumed
\begin{equation}
    b_{1}-b_{1}^{\prime}=b_{1}^{t}\; .
    \label{Eq.b1_redef}
\end{equation}
On the other hand, the ab-initio calculated $b_{2}^{t}$ \DL(no musis nejak uvest, ze ten superscript t ...theoreteical and e je experimental...) constant~\cite{MAELAS_2_r22} refers to the $b_{2}$, as both definitions (Eqs.~\ref{Eq:MagelEng} and~\ref{Eq:2nd_ME}) leads to the same result of the energy difference
\begin{align}
    & \frac{1}{V_{0}}\Big[ E_{me}^{(\mathrm{II})}([\frac{1}{\sqrt{2}  }\frac{1}{\sqrt{2}} 0],\varepsilon_{xx}) - E_{me}^{(\mathrm{II})}([\frac{-1}{\sqrt{2}  }\frac{1}{\sqrt{2}} 0],\varepsilon_{xx}) \Big] \nonumber = \\
    & \qquad = 2b_{2}\varepsilon_{\mathrm{xy}} \, .
\end{align}

In addition, besides the ab-initio constants, experiment-based constants $b_{1}^{e}$, $b_{2}^{e}$ ~\cite{MAELAS_2_r22}, estimated through Eq.~\ref{Eq.lambda001111}, are included to compare the modeled results with the experimental evidence.
Actually, one compares dependencies of the equilibrium strains $\varepsilon_{xx}$ and $\varepsilon_{xy}$ as functions of the magnetization direction $\bm{\alpha}$. The equilibrium strain related to the original formula of the magnetoelastic energy, concerning the $b_{1}^{t}$, $b_{2}^{t}$ (Fig.~\ref{fig:bcc_fcc} dash-dotted lines), resp. $b_{1}^{e}$, $b_{2}^{e}$ (Fig.~\ref{fig:bcc_fcc} dotted lines) constants,  is evaluated based on the analytical expression (Eqs.~\ref{Eq.equilb_strain_ii},~\ref{Eq.equilb_strain_ij}), where the term including $b_{0}$ is neglected due to the comparison. Whereas, for the extended magnetoelastic energy formula (Eq.~\ref{Eq:2nd_ME}), the minimization of the sum of the magnetoelastic and elastic energy is performed for each  $\bm{\alpha}$ direction separately, substituting $\bm{\alpha}$, $b$, and $C_{ij}$ values.

Regarding the modeled bcc Fe and fcc Ni systems, the extended magnetoelastic energy formula brings a constant shift of the $\varepsilon_{xx}$ curve (Fig.~\ref{fig:bcc_fcc}). It brings the theoretical equilibrium strain $\varepsilon_{xx}$ closer to the experimental one at $\bm{\alpha}=[100]$, where only the $b_{1}$ constant plays a role.  The shift arises from the occurrence of the extra linear term $b_{1}^{\prime}\sum_{i\neq j} \varepsilon_{ii}\alpha_{j}$ (Eq.~\ref{Eq:2nd_ME}). Unlike  the full expression of magnetoelastic energy in Eq.~\ref{Eq:2nd_ME}, the analytical solution of the equilibrium strain (Eq.~\ref{Eq:Eminimiz}) is feasible assuming only the linear terms related to the $b_{1}$ ,$b_{2}$, and $b_{1}^{\prime}$ constants.  It reads 
\begin{align}
   & \tilde{\varepsilon}^{{eq}}_{ii}= \frac{(-b_{1}+b_{1}^{\prime})\alpha_{i}^{2}}{C_{11}-C_{12}} 
   + \frac{b_{1}C_{12}-b_{1}^{\prime}C_{11}}{(C_{11}-C_{12})(C_{11}+2C_{12})} \label{Eq.equilb_strain_ii_b3} \; , \\
   &  \tilde{\varepsilon}^{{eq}}_{ij} = \varepsilon^{{eq}}_{ij}= -\frac{b_{2}\alpha_{i}\alpha_{j}}{2C_{44}} , i\neq j \;  \label{Eq.equilb_strain_ij_b3}  \; ,
\end{align}
which leads to similarly  modified  magnetostriction coefficients
\begin{align}
  & { \tilde{\lambda}^{\alpha} = \frac{-b_{0}+(-b_{1}+b_{1}^{\prime})/3}{C_{11}+2C_{12}} }\\
  & \tilde{\lambda}_{001}=\frac{2(-b_{1}+b_{1}^{\prime})}{3(C_{11}-C_{12})}  \\
  & \tilde{\lambda}_{111}=\lambda_{111}= \frac{-b_{2}}{3C_{44}} 
\end{align}

Assuming the original equilibrium strain formula (Eq.~\ref{Eq.equilb_strain_ii}), the magnetization direction independent term contributing to the volume magnetostriction is modified (Eq.~\ref{Eq.equilb_strain_ii_b3}), which is responsible for the constant shift (Fig.~\ref{fig:bcc_fcc} dashed lines vs. dash-dotted ones). 

Comparing the behavior given only by the linear terms (Eqs.~\ref{Eq.equilb_strain_ii_b3},~\ref{Eq.equilb_strain_ij_b3}) (Fig.~\ref{fig:bcc_fcc} dashed lines )  with the equilibrium strain related to magnetoelastic energy higher-order strain terms (Fig.~\ref{fig:bcc_fcc} solid), it is evident that the quadratic terms do not noticeably modify the results. Regarding the derivation arising from the dipole-dipole interaction (Eq.~\ref{Eq:2nd_ME}), the estimated model values of $b^{\prime\prime}$ constants (Table~\ref{Tab:magel_constants}), related to the quadratic strain terms, are small compared to the elastic constants appearing in the similar strain terms in the equilibrium strain condition. It leads to negligible influence on the higher-order terms. They might be more important for systems with lower symmetry, i.e., tetragonal ones, where the magnetoelastic effects are stronger~\cite{Nieves_sss_r25_seconorder}.

 \FloatBarrier
\section{Conclusions}

In this work, assuming bcc and fcc structures, we discussed the impact of the higher-order strain terms in the magnetoelastic energy on the magnetostriction behavior. Considering the dipole-dipole interaction, we derived higher-order terms of the magnetoelastic energy and evaluated the magnetostriction behavior considering the model magnetoelastic constant.  
It shows that the higher  strain terms in the magnetoelastic energy under the cubic symmetry do not play a significant role from the points of view of the anisotropic magnetostriction, as the modeled magnetoelastic constants related to the quadratic terms are small. {It also means that the opposite signs of the Fe constant $b_{2}$ given by the ab-initio calculations resp. experiments, do not originate in the missing high order corrections.}
On the other hand, the isotropic magnetostriction can be substantially modified by the occurrence of an extra linear term, which is not included in the common definition of the magnetoelastic energy.

\FloatBarrier
\section*{Declaration of Interest Statement}
The authors declare that they have no known competing financial interests or personal relationships that could have appeared to influence the work reported in this paper.

\section*{Author contributions}

\textbf{JŠ}: Conceptualization, Data curation , Formal analysis, Investigation, Visualization, Writing – original draft;  \textbf{OF}: Data curation, Investigation,  Writing – review \& editing;  \textbf{DL}: Formal analysis, Writing – review \& editing

\section{Acknowledgemets}
{JS and OF acknowledges the GAČR project  24-11388I of the Grant Agency of Czech Republic and DL the QM4ST project No. CZ.02.01.01/00/22\_008/0004572  by the Ministry of Education, Youth
and Sports of the Czech Republic. Computational resources were provided by the e-INFRA CZ project (ID:90254), supported by the Ministry of Education, Youth and Sports of the Czech Republic.}

\onecolumn
\clearpage
\appendix*

\section{High order expansion details}
\label{APPENDIX}


To evaluate the strain-dependent terms $\Delta E_{dd}$ (Eq.~\ref{Eq:Edd_difference}), one has to expand the differences in the direction cosines $\tilde{\beta}_{i}$ (Eq.~\ref{Eq.dircosmod}) and the bond length $r$ (Eq.~\ref{Eq.raddist_mod}).
Starting from the term
\begin{equation}
    \frac{\partial l}{\partial r}\Big\vert_{r_{0}}  \Delta {r} \Bigg[ (\bm{\alpha}\cdot\tilde{\bm{\beta}})^{2}  - \frac{1}{3} \Bigg] \; ,
    \label{Eq.Edd_term1}
\end{equation}
a Taylor series  expansion is used for the bond length $r$ as follows
    \begin{align}
    &r^{\prime} = r_{0} \;  \Bigg\vert 1+\frac{\bm{\varepsilon}\mathbf{r}_{0}}{r_0^2} \Bigg\vert =  \\
    &= r_{0} \Big[  1 +\frac{1}{r_{0}^2}  \Big( (\varepsilon_{xx}r_{x}+\varepsilon_{xy}r_{y}+\varepsilon_{xz}r_{z})^{2} 
    +(\varepsilon_{yx}r_{x}+\varepsilon_{yy}r_{y}+\varepsilon_{yz}r_{z})^{2}  
    + (\varepsilon_{xx}r_{x}+\varepsilon_{xy}r_{y}+\varepsilon_{xz}r_{z})^{2}  \Big) \Big]^{\frac{1}{2}}  = \nonumber \\
    & = r_{0} \sqrt{1 + t} \approx  r_{0} \Big( 1 +\frac{t}{2} - \frac{t^2}{8} +\frac{t^3}{16} + \mathcal{O}(t^4) \Big) ; 
    \qquad t = \frac{1}{r_{0}^2} \sum_{i} \Big[ \sum_{j} \varepsilon_{ij} r_{j} \Big]^2 ; i,j =x,y,z \; . \nonumber
    \end{align}

Then, the approximate bond length difference reads
\begin{align}
    &\Delta r \sim r_{0}\Big[ \frac{t}{2} - \frac{t^2}{8} +\frac{t^3}{16} \Big]; \qquad t = \frac{1}{r_0^2} \sum_{i} \Big[ \sum_{j} \varepsilon_{ij} r_{j} \Big]^2 ; i,j =x,y,z \; .
\end{align}
To evaluate  term Eq.~\ref{Eq.Edd_term1}, contributions belonging to the interaction with the neighboring atoms have to be summed up. Here, only the nearest-neighbor contributions are considered.

The next term
\begin{equation}
    l(r_{0})(\bm{\alpha}\cdot\Delta\tilde{\bm{\beta}})^{2}
\end{equation}
is related to the modification of the direction cosine $\tilde{\beta}$
Assuming a strained direction cosine $\tilde{\beta}^\prime$ 
a term containing $\bm{\alpha}\tilde{\bm{\beta}}^{\prime}$ can be expanded as 
follows
\begin{align}
    &l(r) \Big[  \bm{\alpha}\tilde{\bm{\beta}}^{\prime}  \Big] = 
     l(r)  \Big[ \frac{\alpha_{x}}{\sqrt{n}} \tilde{\beta}_{x}^{\prime}  \quad  + \frac{\alpha_{y}}{\sqrt{n}}   \tilde{\beta}_{y}^{\prime} \quad + \frac{\alpha_{z}}{\sqrt{n}} \tilde{\beta}_{z}^{\prime}  \Big]^{2}    \\
&= l(r) \Big[ \frac{\alpha_{1}}{\sqrt{n}} \Big( X +A \Big) + \frac{\alpha_{2}}{\sqrt{n}}  \Big( Y +B \Big)
  + \frac{\alpha_{3}}{\sqrt{n}} \Big( Z+C  \Big)  \Big]^{2} \nonumber \\
%
        &=  \frac{l(r_{0})}{n}  \Big[ X{^{2}}\alpha_{1}^{2}  + Y{^{2}}\alpha_{2}^{2}   + Z{^{2}}\alpha_{3}^{2} 
        + 2XY\alpha_{1}\alpha_{2}  + 2XZ\alpha_{1}\alpha_{3} + 2YZ\alpha_{2} \alpha_{3} \Big]   \nonumber  \\ 
        & + \frac{l(r_{0})}{n}\Bigg[ {  \Big [  2X\alpha_{1}\alpha_{1} + 2Y\alpha_{1}\alpha_{2}   + 2Z\alpha_{1}\alpha_{3}  \Big]A} 
        {+ \Big [   2X\alpha_{1}\alpha_{2} + 2Y\alpha_{2} \alpha_{2}  + 2Z\alpha_{2}\alpha_{3}\Big]  B }   \nonumber     \\  &
        \qquad \qquad{+ \Big [ 2X\alpha_{1}\alpha_{3} +2Y\alpha_{2} \alpha_{3}   + 2Z\alpha_{3} \alpha_{3}   
   \Big] C} \nonumber\\
        &  \qquad \qquad+  \Big [  \alpha_{1}^{2}A^{2}   + \alpha_{2}^{2}B^{2}
 +\alpha_{3}^{2}C^{2}  \Big] 
 +   \Big[ 2\alpha_{1}\alpha_{2}AB
   +2\alpha_{1}\alpha_{3}AC  +2\alpha_{2}\alpha_{3}BC  \Big] \Bigg]  \nonumber\\
   & = l(r_{0})\Bigg[  (\bm{\alpha}\tilde{\bm{\beta}})^{2}_{0} +  (\bm{\alpha}\Delta\tilde{\bm{\beta}})^{2}    \Bigg] \; ; 
   \qquad \tilde{\beta_{1}}=X, \tilde{\beta_{2}}=Y, \tilde{\beta_{3}}=Z \quad  \Delta\tilde{\beta_{1}}=A, \Delta\tilde{\beta_{2}}=B, \Delta\tilde{\beta_{3}}=C  \nonumber \;
    \end{align}
where $\sqrt{n}$ denotes the norm of the nearest neighbor bond vector. Summing the terms containing $A$, $B$, $C$, defined by Eqs.~\ref{Eq.:dircoschange},~\ref{Eq.dircosmod}, over the nearest neighbors a contribution related to $\Delta(\bm{\alpha}\tilde{\bm{\beta}})^{2}$ in Eq.~\ref{Eq:Edd_difference} can be calculated. 
To obtained a  change of the direction cosine $\Delta \tilde{\beta}_{i}$,  the strained direction cosines $\tilde{\beta}_{i}^{\prime}$ (Eq.~\ref{Eq.:dircoschange}) are expanded using Taylor series of $1/\sqrt{1+x}$ as follows
\begin{align}
    &\tilde{\beta}_{i}^{\prime} = \frac{r_{i}^{\prime}}{ r_{0} } \frac{1}{\vert 1+(\bm{\varepsilon}\mathbf{r}_{0})/r_{0}^2 \vert}  \label{Eq.:dircos_expa} 
    =\frac{r_{i}+\sum_{j} \varepsilon_{ij}r_{j}}{ r_{0} } \Big(1+\frac{1}{r_{0}^2} \sum_{i} \Big[ \sum_{j} \varepsilon_{ij} r_{j}\Big]^2 \Big)^{-\frac{1}{2}}  \\
    &=\frac{r_{i}+\sum_{j} \varepsilon_{ij}r_{j}}{ r_{0} } \Big(1+t\Big)^{-\frac{1}{2}}  
    \approx \frac{r_{i}+\sum_{j} \varepsilon_{ij}r_{j}}{ r_{0} } \Big[ 1 -\frac{t}{2} + \frac{3t^2}{8} -\frac{5t^3}{16} + \mathcal{O}(t^4) \Big] \; , \nonumber
\end{align}
which leads to the difference in the direction cosine
\begin{equation}
    \Delta \tilde{\beta}_{i}\sim \frac{r_{i}+\sum_{j} \varepsilon_{ij}r_{j}}{ r_{0} } \Big[ 1 -\frac{t}{2} + \frac{3t^2}{8} -\frac{5t^3}{16}  \Big] - \frac{r_{i}}{r_{0}}  \; .
    \label{Eq.:dircos_diff1}
\end{equation}
The last  extra term
\begin{equation}
     \frac{\partial l}{\partial r} \Big\vert_{r_{0}} \Delta {r}  \Bigg[  (\bm{\alpha}\cdot\Delta\tilde{\bm{\beta}})^{2}   \Bigg]
\end{equation}

combines the changes of the bond length and direction cosines described above.
Actually, the strain-dependent contributions $\Delta E_{dd}$ to the dipole-dipole interaction, resulting in the magnetoelastic energy (Eq.~\ref{Eq:2nd_ME}), were evaluated employing the  Python library for symbolic mathematics -- \textit{SymPy}~\cite{SymPy}.
It gives the magnetoelastic energy up to the third order in the strain in the case of the bcc structure as

\begin{align}
   & E_{me}^{III}  =   L_{0} \Bigg[  \frac{32 }{9} \sum_{i}\varepsilon_{ii}\alpha_{i}^{2}   -  \frac{16 }{9}\sum_{i\neq j }\varepsilon_{jj} \alpha_{i}^{2}  \Bigg] +
     \frac{16 }{9}(L_{0} +L_{1} )\sum_{ij} \varepsilon_{ij}  \;  \alpha_{i} \alpha_{j}   \\
      & + \frac{1}{27}\sum_{i\neq j}\Bigg[ -16(L_{0}  -2 L_{1})\varepsilon_{ii}^2 + 8(L_{0}  -2 L_{1})\varepsilon_{jj}^2   -8(5 L_{0}  -4 L_{1})\varepsilon_{ij}^2     -16(2 L_{0}  -L_{1})\varepsilon_{ii} \varepsilon_{jj}    \Bigg] \alpha_{i}^{2} \nonumber \\
         & +  \frac{1}{27}\sum_{i\neq j \neq k \neq i}\Bigg[ + 8(5 L_{0}  -4 L_{1})\varepsilon_{jk}^2  + 16(2 L_{0}  -L_{1})\varepsilon_{jj} \varepsilon_{kk}   \Bigg] \alpha_{i}^{2} \nonumber \\
   &  +\frac{1}{27}\sum_{i\neq j}\Bigg[ -80(L_{0}  -L_{1})\varepsilon_{ii} \varepsilon_{ij} \Bigg] + \frac{1}{27}\sum_{i\neq j\neq k \neq i}\Bigg[  + 32(L_{0}  -L_{1})\varepsilon_{ij} \varepsilon_{kk} + 8(11 L_{0}  -5 L_{1})\varepsilon_{ik} \varepsilon_{jk}
   \Bigg]  \alpha_{i} \alpha_{j} \nonumber  \\
      &+ \frac{1}{27}\sum_{i\neq j}\Bigg[\frac{1}{3}(-64L_0 + 16L_1)\varepsilon_{ii}^3 + \frac{1}{3}(32L_0-8L_1)\varepsilon_{jj}^3 
         +  32(L_0-L_1)\varepsilon_{jj}\varepsilon_{ii}^2 +(-16L_0+8L_0)\varepsilon_{ii}\varepsilon_{jj}^2 \nonumber \\
         &+
        (16L_0-40L_1)\varepsilon_{ij}^2\varepsilon_{ii} +  (16L_0-8L_1)3\varepsilon_{ij}^2\varepsilon_{jj}
        \Bigg]\alpha_i^2  \nonumber\\
    &+\frac{1}{27} \sum_{i\neq j \neq k \neq i}\Bigg[(-32L_0+24L_1)\varepsilon_{jj}\varepsilon_{kk} + (-16L_0+8L_1)\varepsilon_{jk}^2\varepsilon_{ii} 
    + (16L_0-8L_1)\varepsilon_{ij}^2\varepsilon_{kk} +(-64L_0+64L_1)\varepsilon_{jk}^2\varepsilon_{jj}    \Bigg]\alpha_i^2  \nonumber \\
     &+  \frac{1}{27}\sum_{i\neq j}\Bigg[\frac{1}{3}(64L_0-80L_1)\varepsilon_{ij}^3 + (96L_0-80L_1)\varepsilon_{ii}^2\varepsilon_{ij} - 32L_0\varepsilon_{ii}\varepsilon_{jj}\varepsilon_{ij}\Bigg]\alpha_i\alpha_j  \nonumber \\
        &      \frac{1}{27}\sum_{i\neq j \neq k \neq i}\Bigg[(-80L_0+64L_1)\varepsilon_{kk}\varepsilon_{ik}\varepsilon_{jk} +(-96L_0+64L_1)\varepsilon_{ii}\varepsilon_{ik}\varepsilon_{jk} +  32L_0\varepsilon_{jj}\varepsilon_{kk}\varepsilon_{ij}  \nonumber
    \\
   &  +  (-48L_0+32L_1)\varepsilon_{kk}^2\varepsilon_{ij} +  (32L_0-16L_1)\varepsilon_{ik}^2\varepsilon_{ij}\Bigg]\alpha_i\alpha_j
     \quad i,j,k \in \{x,y,z \} \nonumber
\end{align}

where for the fcc symmetry it yields
    \begin{align}
        & E_{me}^{III}  =  \sum_{i}{ (4L_{0}+2L_{1}) \varepsilon_{ii}} \alpha_{i}^{2} +    \sum_{i\neq j}{ (-2L_{0}+L_{1}) \varepsilon_{jj}} \alpha_{i}^{2}   
       + \frac{1}{2}(8L_{0}+4L_{1}) \sum_{i \neq j}\varepsilon_{ij}  \;  \alpha_{i} \alpha_{j}   \\
      & + \frac{1}{4}\sum_{i\neq j}\Bigg[  -2(4 L_{0}  -5 L_{1})\varepsilon_{ii}^2 + (4 L_{0}  -3 L_{1})\varepsilon_{jj}^2  + 2(2 L_{0} + L_{1})\varepsilon_{ij}^2      -2 \varepsilon_{ii} \varepsilon_{jj} L_{1}      \Bigg] \alpha_{i}^{2} \nonumber \\
    & + \frac{1}{4}\sum_{i\neq j \neq k \neq i}\Bigg[   -\frac{1}{2}4(2 L_{0}  -L_{1})\varepsilon_{jk}^2   \Bigg] \alpha_{i}^{2} \nonumber \\
     & + \sum_{i \neq j}\Bigg[2 \varepsilon_{ii} \varepsilon_{ij} L_{1} \Bigg]  \alpha_{i} \alpha_{j} +  \sum_{i \neq j \neq k \neq i}\Bigg[    -\frac{1}{2}4(2 L_{0}  -L_{1})\varepsilon_{ij} \varepsilon_{kk}  -\frac{1}{2}2(6 L_{0}  -5 L_{1})\varepsilon_{ik} \varepsilon_{jk}
   \Bigg]  \alpha_{i} \alpha_{j} \nonumber \\
   & +\frac{3L_0}{4}\sum_{i}\varepsilon_{ii}^3\alpha_{ii}^2 
        +   \sum_{i\neq j}\Bigg[-\frac{3L_0}{8}\varepsilon_{jj}^3+
    (2L_0-\frac{9}{8}L_1)\varepsilon_{jj}\varepsilon_{ii}^2+(-2L_0+\frac{11}{8}L_1)\varepsilon_{jj}^2\varepsilon_{ii} \nonumber \\
    %
    & +  (2L_0-\frac{9}{4}L_1)\varepsilon_{ii}\varepsilon_{ij}^2+(-4L_0+3L_1)\varepsilon_{jj}\varepsilon_{ij}^2 +\Bigg]\alpha_{ii}^2 \nonumber \\
    &  +  \sum_{i\neq j\neq k \neq i}   \Bigg[ -\frac{L_1}{4}\varepsilon_{kk}\varepsilon_{ij}^2 
        -\frac{1}{2}\frac{L_1}{2}\varepsilon_{ii} \varepsilon_{jk}^2
    +(-2L_0+\frac{5}{4}L_1)\varepsilon_{jj} \varepsilon_{jk}^2 
    -  \frac{1}{2}   2\varepsilon_{ij}\varepsilon_{ik}\varepsilon_{jk}L_1\Bigg]\alpha_{ii}^2 \nonumber \\
    %
    %
    & + \sum_{i\neq j} \Bigg[ (2L_0-2L_1) \varepsilon_{ij}\varepsilon_{jj}^2  + \frac{1}{2}(-4L_0+2L_1)\varepsilon_{ij}^3\Bigg ]\alpha_i\alpha_j \nonumber  \\
     & +    \sum_{i\neq j\neq k \neq i} \Bigg[  +   \frac{1}{2}(4L_0-3L_1) \varepsilon_{ij}\varepsilon_{kk}^2 + (2L_0-4L_1)\varepsilon_{ij}\varepsilon_{ik}^2\Bigg]\alpha_i\alpha_j \nonumber \\
        &+\sum_{i\neq j}\Bigg[\frac{1}{2}(-8L_0+5L_1)\varepsilon_{ii}\varepsilon_{jj}\varepsilon_{ij}
    \Bigg]\alpha_i\alpha_j \nonumber \\
    &  +    \sum_{i\neq j\neq k \neq i}\Bigg[(4L_0-5L_1)\varepsilon_{ii}\varepsilon_{ik}\varepsilon_{jk}+\frac{1}{2}(16L_0-12L_1)\varepsilon_{kk}\varepsilon_{ik}\varepsilon_{jk}+
    (4L_0-3L_1)\varepsilon_{ii}\varepsilon_{kk}\varepsilon_{ij}\Bigg]\alpha_i\alpha_j\nonumber  
\end{align}

The equilibrium strain (Eq.~\ref{Eq:Eminimiz}) for the common definition of the magnetoelastic energy (Eq.~\ref{Eq:MagelEng}) can be obtained by solving a simple set of following equations
\begin{align}
   b_{1}\alpha_{x}^2+  C_{11}\varepsilon_{xx} + C_{12}(\varepsilon_{yy}+\varepsilon_{zz} )= 0 \label{Eq:mimim_set_orig_1} \; ,\\
b_{1}\alpha_{y}^2+ C_{11}\varepsilon_{yy} + C_{12}(\varepsilon_{xx}+\varepsilon_{zz}) = 0  \; ,\\
b_{1}\alpha_{z}^2+ C_{11}\varepsilon_{zz} + C_{12}(\varepsilon_{xx}+\varepsilon_{yy}) = 0  \; , \\
2b_{2}\alpha_{1}\alpha_{2}+ 4C_{44}\varepsilon_{xy} = 0  \label{Eq:mimim_set_orig_4}  \; , \\
2b_{2}\alpha_{1}\alpha_{3}+ 4C_{44}\varepsilon_{xz} = 0  \; , \\
2b_{2}\alpha_{2}\alpha_{3}+ 4C_{44}\varepsilon_{yz} = 0  
\label{Eq:mimim_set_orig_6}  \; .
\end{align}

Assuming the extended definition of the magnetoelastic energy with higher order strain terms (Eq.~\ref{Eq:2nd_ME}), the conditions defining the equilibrium strain become strongly coupled. For instance, the Eq.~\ref{Eq:mimim_set_orig_1} changes to 
\begin{align}
&\alpha_{x}^2 b_{1} + b_{1}^{\prime}(\alpha_{y}^2 + \alpha_{z}^2) + 2\alpha_{x}^2 b_{1}^{\prime\prime}\varepsilon_{xx}   + b_{2}^{\prime\prime}(2\alpha_{y}^2 \varepsilon_{xx} + 2\alpha_{z}^2\varepsilon_{xx}) \nonumber\\
& + b_{5}^{\prime\prime}(\alpha_{x}^2 (\varepsilon_{yy} + \varepsilon_{zz}) + \alpha_{y}^2\varepsilon_{yy} + \alpha_{z}^2 \varepsilon_{zz}) + 2b_{6}^{\prime\prime}(\alpha_{y}^2\varepsilon_{zz} + \alpha_{z}^2 \varepsilon_{yy}) \nonumber\\
& + b_{7}^{\prime\prime}(\alpha_{x}\alpha_{y}\varepsilon_{xy}  + \alpha_{x}\alpha_{z} \varepsilon_{xz}) + 2\alpha_{y}\alpha_{z} b_{8}^{\prime\prime}\varepsilon_{yz} \nonumber \\
&  +  C_{11}\varepsilon_{xx} + C_{12}(\varepsilon_{yy}+\varepsilon_{zz} )= 0  \; ,
\label{Eq:minimiz_full1}
\end{align}
 and the Eq.~\ref{Eq:mimim_set_orig_4} is modified to
\begin{align}
  & 2 \alpha_{x} \alpha_{y} b_{2}  + b_{3}^{\prime\prime} (2 \alpha_{x}^2 \varepsilon_{xy} + 2 \alpha_{y}^2 \varepsilon_{xy}) \nonumber \\
  & + 4 \alpha_{z}^2 b_{4}^{\prime\prime} \varepsilon_{xy} +  \alpha_{x} \alpha_{y} b_{7}^{\prime\prime} (\varepsilon_{xx} + \varepsilon_{yy})  \nonumber\\
  &+ 2 \alpha_{x} \alpha_{y} b_{8}^{\prime\prime} \varepsilon_{zz}  + 2 b_{9}^{\prime\prime} (\alpha_{x} \alpha_{z} \varepsilon_{yz} + \alpha_{y} \alpha_{z} \varepsilon_{xz}) + 4C_{44}\varepsilon_{xy} = 0
  \label{Eq:minimiz_full4} \; .
\end{align}

\FloatBarrier

To estimate the  equilibrium strains, the following published values of the magnetoelastic~\cite{MAELAS_2_r22} and elastic constants~\cite{AELAS_r17} stated in the Tables~\ref{Tab:b_parameters} resp.~\ref{Tab:Cij_parameters} were used.

\begin{table}[h]
\centering
\caption{\label{Tab:b_parameters} Ab-initio and experimental value of the Fe and Ni magnetoelstic constants adopted from~\cite{MAELAS_2_r22}}%
\setstretch{1.5}

\begin{tabular}{ccc}
\hline
\hline
 $b$(MPa) &  Fe & Ni   \\
\hline
$b_{1}^{t}$ & -5.9 & 14.4  \\
$b_{2}^{t}$ & -4.9 &  18.5  \\
$b_{1}^{e}$ &  -4.1 &  9.9 \\
$b_{2}^{e}$ &  10.9 &  13.9 \\
\hline
\hline

\end{tabular}

\end{table}

\begin{table}[h]
\centering
\caption{\label{Tab:Cij_parameters} Ab-initio calculated elastic constant adopted from~\cite{MAELAS_1_r21}}%
\setstretch{1.5}

\begin{tabular}{ccc}
\hline
\hline
 $C_{ij}$(GPa) &  Fe & Ni   \\
\hline
$C_{11}$  & 288   & 298  \\
$C_{12}$  & 152  &   166 \\
$C_{44}$  & 104  &   140 \\
\hline
\hline
\end{tabular}

\end{table}

\FloatBarrier

\section{Data availability}
A detailed calculation of high-order magnetoelastic energy terms  is available from the Zenodo repository at  https://doi.org/10.5281/zenodo.17924230

\twocolumn

  \bibliographystyle{elsarticle-num} 
  \bibliography{bibsamp.bib}

@PREAMBLE{
 "\providecommand{\noopsort}[1]{}" 
 # "\providecommand{\singleletter}[1]{#1}%" 
}

@article{SymPy,
author = {Joyner, David and \v{C}ert\'{\i}k, Ond\v{r}ej and Meurer, Aaron and Granger, Brian E.},
title = {Open source computer algebra systems: SymPy},
year = {2012},
issue_date = {September/December 2011},
publisher = {Association for Computing Machinery},
address = {New York, NY, USA},
volume = {45},
number = {3/4},
issn = {1932-2232},
url = {https://doi.org/10.1145/2110170.2110185},
doi = {10.1145/2110170.2110185},
journal = {ACM Commun. Comput. Algebra},
month = jan,
pages = {225–234},
numpages = {10}
}

@article{r25_Ngouagnia_SAW_SW,
    author = {Ngouagnia Yemeli, I. and Christienne, L. and Rovillain, P. and Duquesne, J.-Y. and Anane, A. and Marangolo, M. and Stoeffler, D.},
    title = {A micromagnetic study of surface-acoustic-wave-driven excitation of spin waves in an iron-based conduit},
    journal = {Journal of Applied Physics},
    volume = {137},
    number = {15},
    pages = {153908},
    year = {2025},
    month = {04},
    issn = {0021-8979},
    doi = {10.1063/5.0260092},
    url = {https://doi.org/10.1063/5.0260092},
    eprint = {https://pubs.aip.org/aip/jap/article-pdf/doi/10.1063/5.0260092/20494534/153908_1_5.0260092.pdf},
}

@article{r21_Babu_SAW_SW,
author = {Babu, Nandan K. P. and Trzaskowska, Aleksandra and Graczyk, Piotr and Centała, Grzegorz and Mieszczak, Szymon and Głowiński, Hubert and Zdunek, Miłosz and Mielcarek, Sławomir and Kłos, Jarosław W.},
title = {The Interaction between Surface Acoustic Waves and Spin Waves: The Role of Anisotropy and Spatial Profiles of the Modes},
journal = {Nano Letters},
volume = {21},
number = {2},
pages = {946-951},
year = {2021},
doi = {10.1021/acs.nanolett.0c03692},
    note ={PMID: 33231459},

URL = { 
        https://doi.org/10.1021/acs.nanolett.0c03692
},
eprint = { 
        https://doi.org/10.1021/acs.nanolett.0c03692
}

}

@article{Nieves_sss_r25_seconorder,
title = {Second-order anisotropy due to magnetostriction for {L}10-{FePt}},
journal = {Solid State Sciences},
volume = {160},
pages = {107782},
year = {2025},
issn = {1293-2558},
doi = {https://doi.org/10.1016/j.solidstatesciences.2024.107782},
url = {https://www.sciencedirect.com/science/article/pii/S1293255824003479},
author = {D. Legut and P. Nieves}
}

@article{Ekreem_jmpt_r07,
title = {An overview of magnetostriction, its use and methods to measure these properties},
journal = {Journal of Materials Processing Technology},
volume = {191},
number = {1},
pages = {96-101},
year = {2007},
note = {Advances in Materials and Processing Technologies, July 30th - August 3rd 2006, Las Vegas, Nevada},
issn = {0924-0136},
doi = {https://doi.org/10.1016/j.jmatprotec.2007.03.064},
url = {https://www.sciencedirect.com/science/article/pii/S0924013607002889},
author = {N.B. Ekreem and A.G. Olabi and T. Prescott and A. Rafferty and M.S.J. Hashmi}
}

@Article{Spetzler_scirep_r21,
author={Spetzler, B.
and Bald, C.
and Durdaut, P.
and Reermann, J.
and Kirchhof, C.
and Teplyuk, A.
and Meyners, D.
and Quandt, E.
and H{\"o}ft, M.
and Schmidt, G.
and Faupel, F.},
title={Exchange biased delta-{E} effect enables the detection of low frequency p{T} magnetic fields with simultaneous localization},
journal={Scientific Reports},
year={2021},
month={Mar},
day={05},
volume={11},
number={1},
pages={5269},
issn={2045-2322},
doi={10.1038/s41598-021-84415-2},
url={https://doi.org/10.1038/s41598-021-84415-2}
}

@article{Calkins_jimss_r07,
author = {Frederick T. Calkins and Alison B. Flatau and Marcelo J. Dapino},
title ={Overview of Magnetostrictive Sensor Technology},

journal = {Journal of Intelligent Material Systems and Structures},
volume = {18},
number = {10},
pages = {1057-1066},
year = {2007},
doi = {10.1177/1045389X06072358},
URL = { 
        https://doi.org/10.1177/1045389X06072358
},
eprint = { 
        https://doi.org/10.1177/1045389X06072358
}
}

@article{Bienkovsky_SensAct_r04,
title = {The possibility of utilizing the high permeability magnetic materials in construction of magnetoelastic stress and force sensors},
journal = {Sensors and Actuators A: Physical},
volume = {113},
number = {3},
pages = {270-276},
year = {2004},
note = {New materials and Technologies in Sensor Applications, Proceedings of the European Materials Research Society 2003 - Symposium N},
issn = {0924-4247},
doi = {https://doi.org/10.1016/j.sna.2004.01.010},
url = {https://www.sciencedirect.com/science/article/pii/S0924424704000172},
author = {Adam Bieńkowski and Roman Szewczyk}
}

@article{Kuszewski_iop_r18,
doi = {10.1088/1361-648X/aac152},
url = {https://dx.doi.org/10.1088/1361-648X/aac152},
year = {2018},
month = {may},
publisher = {IOP Publishing},
volume = {30},
number = {24},
pages = {244003},
author = {Kuszewski, P and Camara, I S and Biarrotte, N and Becerra, L and von Bardeleben, J and Savero Torres, W and Lemaître, A and Gourdon, C and Duquesne, J-Y and Thevenard, L},
title = {Resonant magneto-acoustic switching: influence of Rayleigh wave frequency and wavevector},
journal = {Journal of Physics: Condensed Matter}
}

@incollection{CLARK1980531,
title = {Chapter 7 Magnetostrictive rare earth-{Fe2} compounds},
series = {Handbook of Ferromagnetic Materials},
publisher = {Elsevier},
volume = {1},
pages = {531-589},
year = {1980},
issn = {1574-9304},
doi = {https://doi.org/10.1016/S1574-9304(05)80122-1},
url = {https://www.sciencedirect.com/science/article/pii/S1574930405801221},
author = {A.E. Clark}
}

@book{engdahl2000handbook,
  title={Handbook of giant magnetostrictive materials},
  author={Engdahl, G{\"o}ran and Mayergoyz, Isaak D},
  volume={107},
  year={2000},
  publisher={Elsevier}
}

@book{chikazumi2009physics,
  title={Physics of Ferromagnetism},
  author={Chikazumi, S.},
  isbn={9780191569852},
  series={International Series of Monographs on Physics},
  url={https://books.google.cz/books?id=AZVfuxXF2GsC},
  year={2009},
  publisher={OUP Oxford}
}

@book{cullen_book,
  title={Materials Science and Technology},
  author={Cullen, J. and   Clark, A.E. and  Hathaway , K.B.},
  volume={pp. 529},
  year={1994},
  publisher={VCH
Publishings}
}

@article{AELAS_r17,
title = {AELAS: Automatic ELAStic property derivations via high-throughput first-principles computation},
journal = {Computer Physics Communications},
volume = {220},
pages = {403-416},
year = {2017},
issn = {0010-4655},
doi = {https://doi.org/10.1016/j.cpc.2017.07.020},
url = {https://www.sciencedirect.com/science/article/pii/S0010465517302400},
author = {S.H. Zhang and R.F. Zhang}
}

@article{MAELAS_1_r21,
title = {MAELAS: MAgneto-ELAStic properties calculation via computational hig--throughput approach},
journal = {Computer Physics Communications},
volume = {264},
pages = {107964},
year = {2021},
issn = {0010-4655},
doi = {https://doi.org/10.1016/j.cpc.2021.107964},
url = {https://www.sciencedirect.com/science/article/pii/S0010465521000801},
author = {P. Nieves and S. Arapan and S.H. Zhang and A.P. Kadzielawa and R.F. Zhang and D. Legut}
}

@article{MAELAS_2_r22,
title = {MAELAS 2.0: A new version of a computer program for the calculation of magneto-elastic properties},
journal = {Computer Physics Communications},
volume = {271},
pages = {108197},
year = {2022},
issn = {0010-4655},
doi = {https://doi.org/10.1016/j.cpc.2021.108197},
url = {https://www.sciencedirect.com/science/article/pii/S001046552100309X},
author = {P. Nieves and S. Arapan and S.H. Zhang and A.P. Kadzielawa and R.F. Zhang and D. Legut}
}







\end{document}